\documentclass[twocolumn, prl,floatfix,superscriptaddress]{revtex4}
\usepackage{graphicx, subfigure, amsfonts,amssymb,amsmath, array, gensymb,fontenc,color, lipsum}
\usepackage{braket}
\usepackage{tikz}
\newcommand*\circled[1]{\tikz[baseline=(char.base)]{
            \node[shape=circle,draw,inner sep=1pt] (char) {#1};}}
\newcommand{\sgn}{\operatorname{sgn}}
\begin{document}

\title{Inducing Magnetic Phase Transitions in Monolayer CrI$_3$ \emph{via} Lattice Deformations}

\author{Michele Pizzochero}
\email{michele.pizzochero@epfl.ch}
\author{Oleg V.\ Yazyev}
\affiliation{Institute of Physics, Ecole Polytechnique F\'ed\'erale de Lausanne (EPFL), CH-1015 Lausanne, Switzerland}

\affiliation{National Centre for Computational Design and Discovery of Novel Materials (MARVEL), Ecole Polytechnique F\'ed\'erale de Lausanne (EPFL), CH-1015 Lausanne, Switzerland}

\date{\today}

\begin{abstract}
 Atomically thin films of layered  chromium triiodide (CrI$_3$) have recently been regarded as suitable candidates to a wide spectrum of technologically relevant applications, mainly owing to the opportunity they offer to achieve a reversible transition between  coexisting in-plane ferro- and  out-of-plane antiferro-magnetic orders. However,  no routes for inducing such a transition have been designed down to the single-layer limit. Here, we address the magnetic response of monolayer CrI$_3$  to in-plane lattice deformations through a combination of isotropic Heisenberg spin Hamiltonians and first-principles calculations. Depending on the magnitude and orientation of the lattice strain exerted, we unveil a series of direction-dependent parallel-to-antiparallel spins crossovers, which yield the emergence of ferromagnetic, N\'eel antiferromagnetic, zigzag and stripy antiferromagnetic ground states. Additionally, we identify a critical point in the magnetic phase diagram whereby the exchange couplings vanish and the magnetism is quenched. Our work establishes guidelines for extensively tailoring the spin interactions in monolayer CrI$_3$ \emph{via} strain engineering, and further expands the magnetically ordered phases which can be hosted in a two-dimensional crystal.
\end{abstract}

\maketitle
\paragraph{Introduction.} 
Over a decade after the first isolation of graphene \cite{Novoselov04}, the class of two-dimensional materials currently encompasses a broad range of diverse properties \cite{Novoselov05, Nicolosi13, Mounet18, CastroNeto09, Yazyev15, Chen2018, Pedramrazi19}. Among them, the emergence of correlated electron phases, including superconductivity, Mott-insulating phases, or charge density waves \cite{CastroNeto01, Cao18a, Cao18b, Manzeli17, Pizzochero19bis}, has culminated in the discovery of intrinsic magnetism in certain atomically thin crystals \cite{Burch2018, Gong19}. In these systems, the pivotal role played by the single-ion anisotropies preserves magnetic order in low dimensions \cite{Mermin66}, as unambiguously demonstrated by the groundbreaking exfoliation of the bulk ferromagnet CrI$_3$ down to the monolayer limit \cite{Huang2017}, rapidly followed by the appearance of many other ultrathin magnets, \emph{e.g.}\ Fe$_3$GeTe$_2$, Cr$_2$Ge$_2$Te$_6$, FePSe$_3$ \cite{Gibertini19, Torelli2019, Burch2018, Gong19, Avsar19}

Within the constantly expanding family of two-dimensional magnets, semiconducting thin films of CrI$_3$ are arguably the most prototypical members. They exhibit a layered crystal structure, in which Cr atoms are located at the corners of in-plane honeycomb networks and reside in edge-sharing octahedra formed by I atoms \cite{McGuire15}. As far as the magnetism is concerned, each Cr$^{3+}$ ion  features effective spin $3/2$ \cite{Lado17}, which interacts with neighboring sites through intra-layer (inter-layer) ferromagnetic (antiferromagnetic) exchange couplings, as pinpointed by the evolution of the thickness-dependent hysteretic features reported in recent MOKE experiments \cite{Huang2017}. The competition between ferro- and  antiferro-magnetic couplings in bi- and few-layer CrI$_3$ allows one to switch between these two phases, thereby tailoring the nature of the magnetic ground state. This has been experimentally accomplished through a broad variety of different techniques, including electrostatic gating \cite{Huang2018}, doping \cite{Jiang2018}, applied external pressure \cite{Song19, Tingxin19}, off-plane lattice deformations \cite{Thiel19}, or adjusting the stacking patterns of the layers \cite{Sivadas18, Ubrig19}. Such an extensive control over the magnetism in atomically thin CrI$_3$ holds promise for the realization of novel ultrathin magnetoelectronic devices \cite{Wang2018, Song1214, Klein1218}.

Down to the single-layer limit, however, the lack of antiferromagnetic interactions across the crystal hinders the realization of any magnetic order other than the ferromagnetic one.  In this Letter, we theoretically investigate the effect of  in-plane lattice deformations on the magnetism of monolayer CrI$_3$. We unveil a series of magnetic phase transitions arising from the competition between symmetry inequivalent Heisenberg exchange couplings in the distorted lattice, which gives rise to both antiferromagnetic and ferrimagnetic orders, along with a non-magnetic phase. Overall, our findings provide useful guidelines for modulating the magnetic properties of CrI$_3$ down to the ultimate limit of atomic thickness through strain engineering.

\medskip
\paragraph{Methodology.} 
Our first-principles calculations are performed within the spin-polarized Density Functional Theory (DFT) framework, as implemented in \textsc{vasp} \cite{Kresse93, Kresse96}. We adopt  the generalized gradient approximation devised by Perdew, Burke and Ernzerhof \cite{PBE}, and further include a Hubbard correction $U = 1.5$ eV on the $d$ orbitals of Cr atoms \cite{DFTU}. Such an exchange-correlation density functional ensures an excellent description of magnetic properties of monolayer CrI$_3$ when compared to benchmark multireference wavefunction results \cite{Pizzochero19} and available experimental data \cite{Huang2017, Torelli18}. The cutoff on the kinetic energy is set to 400 eV and the Brillouin zone is sampled  with a 12 $\times$ 12 $k$-mesh. We determine the exchange couplings between a pair of magnetic Cr$^{3+}$ sites through the well-established four-state method for energy mapping analysis detailed in Refs.\ \cite{Xiang11, Xiang13}.  This approach allows one to determine the exchange coupling between a pair of sites without introducing any arbitrary choice on the magnetic configurations of the supercells. For each spin configuration, we optimized the atomic coordinates until the residual Hellmann-Feynman forces acting on each atom  converge to 0.005 meV/{\AA}. In order to properly account for all the relevant exchange couplings between the nearest neighbors in the distorted lattice, we consider an atomic model of monolayer CrI$_3$ consisting of a 2 $\times$ 2 hexagonal supercell containing 32 atoms.  For the calculation of the exchange couplings between the second- and third-nearest neighbors, on the other hand, we adopt a 3 $\times$ 3 hexagonal supercell containing 72 atoms.

   \begin{figure}[]
  \centering
  \includegraphics[width=1\columnwidth]{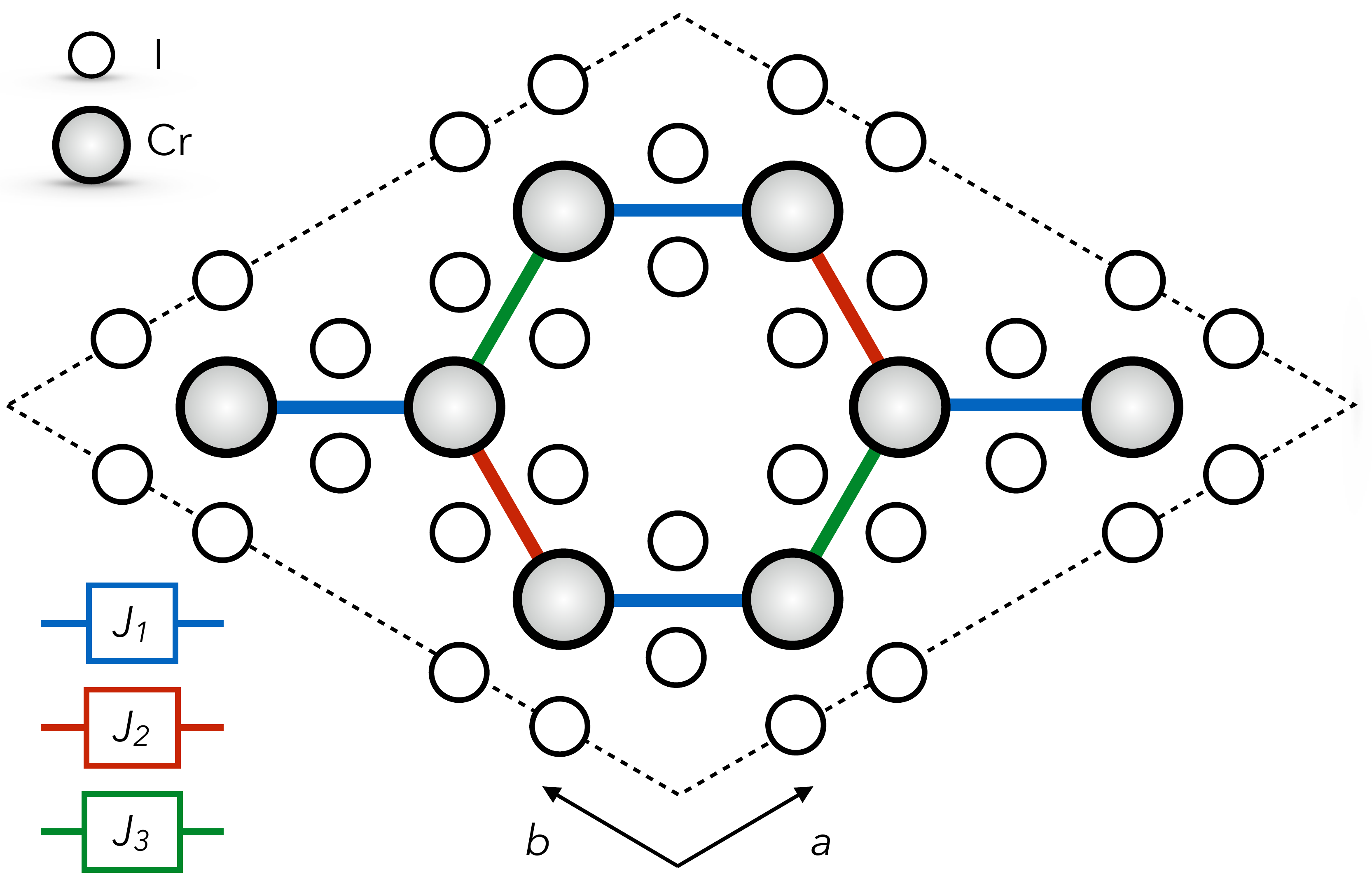}
  \caption{Atomic structure of monolayer CrI$_3$, with white and grey circles representing chromium and iodine atoms, respectively. The 2 $\times$ 2 supercell considered in our calculations is indicated with black dashed lines. The three symmetry inequivalent exchange couplings $J_1$, $J_2$, and $J_3$ between Cr$^{3+}$ ions are indicated with blue, red, and green solid lines, respectively.} \label{Fig1}
\end{figure}

\medskip
\paragraph{Results and Discussion.} 
We start by describing magnetic interactions in unstrained monolayer CrI$_3$. According to our previous quantum chemistry investigation \cite{Pizzochero19}, the isotropic exchange coupling exceeds inter-site anisotropies in CrI$_3$ by several orders of magnitude. This indicates that the \emph{isotropic} Heisenberg Hamiltonian $\mathcal{H}$ which couples the $i$-th and $j$-th  sites carrying $S_i$ and $S_j$ spins, respectively, is a suitable spin model for the description of the magnetic exchange interactions in monolayer CrI$_3$ \cite{Lado17, Pizzochero19}. Such a Hamiltonian reads as 

\begin{equation}
 \mathcal{H}  = \frac{J}{2}\sum_{\braket{i,j}} \vec{S_i} \cdot \vec{S}_{j} + \frac{J'}{2}\sum_{\braket{\braket{i,j}}} \vec{S_j} \cdot \vec{S}_{j} + \frac{J''}{2} \sum_{\braket { \braket { \braket {i,j}}}} \vec{S_i} \cdot \vec{S}_{j}
 \label{Eq1}
\end{equation}

with $J$, $J'$ and $J''$ being the Heisenberg exchange couplings between nearest, second-nearest, and third-nearest neighbors sites, respectively. We determine $J$, $J'$ and $J''$ to be $-1.53$ meV, $-0.38$ meV, and $-0.01$ meV. In line with earlier calculations \cite{Torelli18}, we note that the inter-site interactions are primarily dictated by the nearest neighbors exchange coupling $J$, as the next-nearest neighbors couplings $J'$ and $J''$ are substantially weaker and further feature the same sign, thereby ruling out any possible competition between $J$, $J'$, and $J''$ in determining the magnetic ground state. Hence, in the following we restrict our description to nearest neighbors couplings only ($J' = J'' = 0$ meV), according to which  Eqn.\ (\ref{Eq1}) boils down to

\begin{equation}
\mathcal{H}  = \frac{J}{2}\sum_{\braket{i,j}} \vec{S_i} \cdot \vec{S}_{j} 
\end{equation}

 \begin{figure}[]
    \includegraphics[width=1.0\columnwidth]{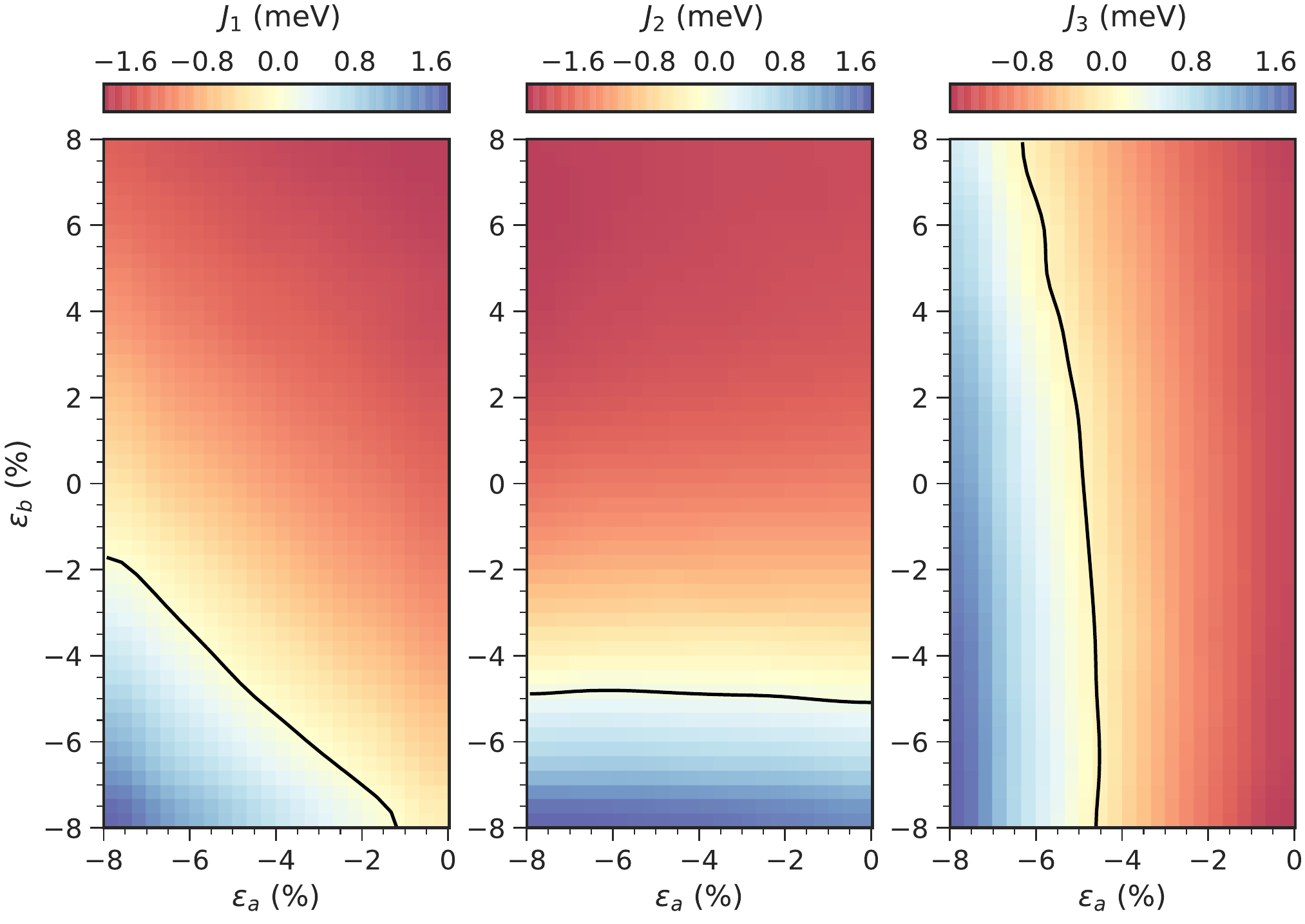}
  \caption{Evolution of $J_1$ (left panel),  $J_2$ (middle panel), and $J_3$ (right panel) with the in-plane $\epsilon_a$ and $\epsilon_b$ lattice deformations in monolayer CrI$_3$. The exchange couplings are defined according to Fig.\ \ref{Fig1}. The black lines indicate the boundaries between positive and negative couplings, corresponding to bond-dependent parallel-to-antiparallel spin transitions. The raw data (given in the Supplementary Information) have been smoothed with cubic polynomials for visualization purposes.} \label{Fig2}
  \end{figure}
  
We systematically investigate the effect of lattice deformations on single-layer CrI$_3$ by introducing in-plane lattice strain along both the $a$ and $b$ directions shown in Fig.\ \ref{Fig1}. We quantify the relative amount of strain along each direction as $\epsilon_a = (a-a_0)/a_0$ and $\epsilon_b = (b-b_0)/b_0$, with $a_0=b_0$ being the experimental equilibrium lattice constant of 13.734 {\AA} in the adopted $2 \times 2$ supercell. We span an interval of lattice strain $ \epsilon_a = [-8\%, 0]$ and  $ \epsilon_b = [-8\%, +8\%]$ through a 9 $\times$ 17 grid. Upon the introduction of an anisotropic strain (that is, $\epsilon_a \neq \epsilon_b$), the honeycomb lattice on which the Cr$^{3+}$ ions reside is subjected to a non-symmetric distortion, which gives rise to the formation of the three inequivalent bonds between the nearest spin-$3/2$ centers. Such a symmetry breaking, in turn, requires the introduction of three distinct Heisenberg exchange couplings $J_1$, $J_2$, and $J_3$ (see Fig.\ \ref{Fig1}) in order to establish the nature of the magnetic ground state. It is clear that these bond-dependent couplings become equivalent by symmetry ($J_1 = J_2 = J_3$) in either the unstrained  ($\epsilon_a = \epsilon_b = 0$) or biaxially strained ($\epsilon_a = \epsilon_b \neq 0$) lattice. Overall, the determination of $J_1$, $J_2$, and $J_3$ within the strain grid mentioned above requires over 1800 calculations.
  
The response of $J_1$, $J_2$, and  $J_3$ to lattice deformations is presented in Fig.\ \ref{Fig2}. While the strain exerted does not affect the of the local magnetic moments carried by the Cr$^{3+}$ ions, it is found to have a strong influence on the magnitude and sign of the exchange couplings. On the one hand, tensile strain lowers all of them, hence further strengthening the ferromagnetic phase. On the other hand, lattice compressions are found to increase $J_1$, $J_2$, and  $J_3$ and eventually revert their signs, thereby promoting a bond-dependent parallel-to-antiparallel spins transition. Indeed, the amount and direction of the lattice strain which is necessary to promote such a transition largely differ for the three couplings. Specifically, our calculations indicate that  an antiparallel orientation of the spins along each pair of Cr$^{3+}$ ions occurs at ($\epsilon_a \lesssim -1.5 \%$; $\epsilon_b \lesssim -1.5 \%$),  ($\epsilon_b \lesssim  -5 \%$), and ($\epsilon_a \lesssim-4.5 \%$) for $J_1$, $J_2$, and  $J_3$, respectively.

   \begin{figure}[]
   \centering
    \includegraphics[width=1.0\columnwidth]{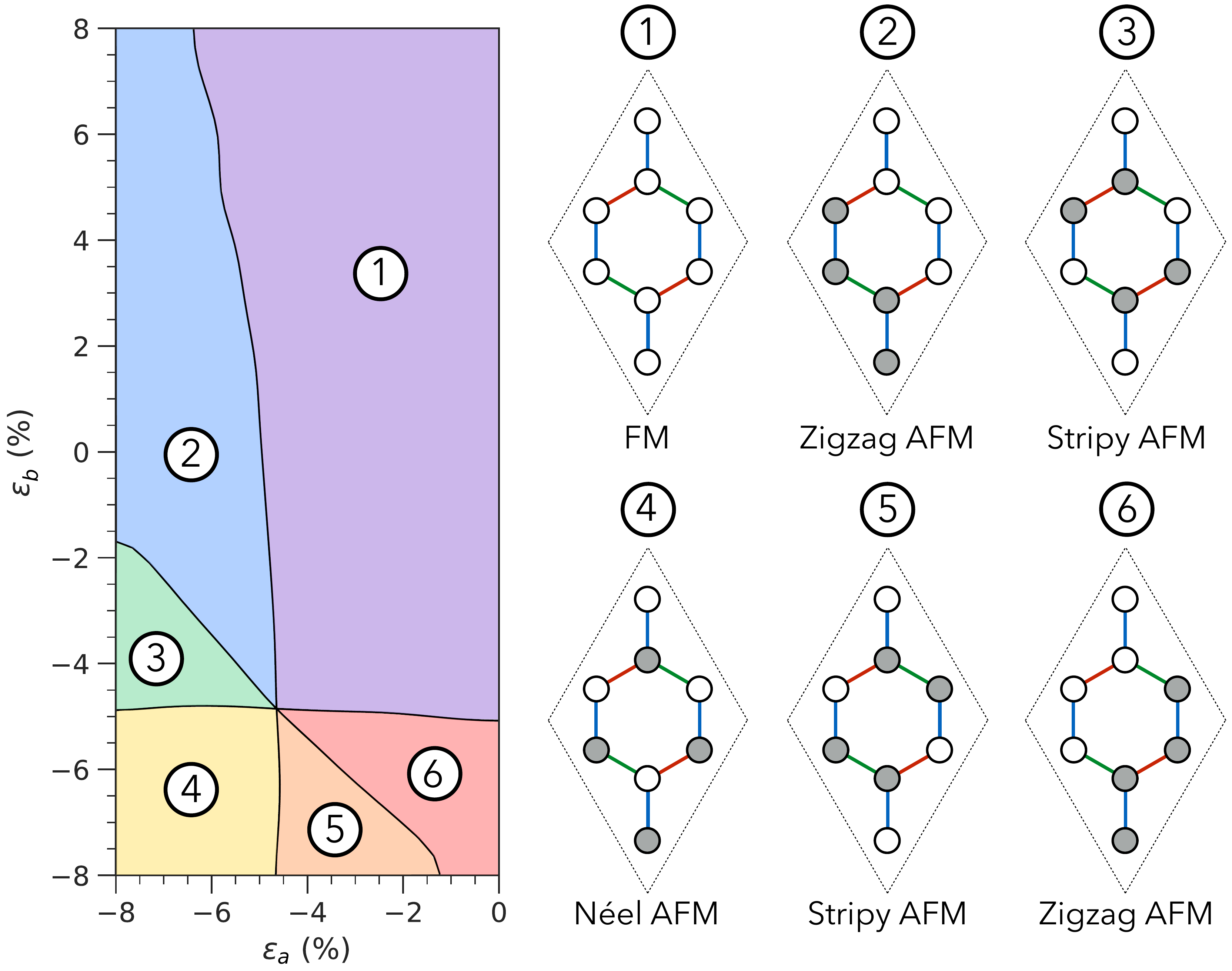}
  \caption{Magnetic phase diagram of monolayer CrI$_3$ under in-plane lattice deformations. The spin configurations emerging in the honeycomb network are also shown, with white and black circles representing Cr$^{3+}$ ions carrying opposite magnetic moments. Bonds between Cr$^{3+}$ ions are color-coded according to Fig.\ \ref{Fig1}.  Dashed lines indicate the unit cell}.
  \label{Fig3}
\end{figure}

Such a pronounced  competition between the three Heisenberg exchange couplings induces the emergence of a strain-dependent magnetic ground state in monolayer CrI$_3$.  Indeed, six distinct domains can be distinguished in the space spanned by the lattice strains, depending on the (relative) signs of  $J_1$, $J_2$, and  $J_3$. The magnetic phase diagram of strained CrI$_3$, along with the spin configurations which are realized upon  lattice deformations, are shown Fig.\ \ref{Fig3}. In Domains \circled{1} and \circled{4}, all the exchange couplings feature the same sign, yielding a ferromagnetic or N\'eel antiferromagnetic ground state, respectively, whereas the other regions exhibit a ferrimagnetic ground state whereby the equality $\sgn(J_1) = \sgn(J_2)  = \sgn(J_3)$ does not hold.  Specifically, Domains \circled{2} and \circled{6} enclose the zigzag antiferromagnetic state, where two exchange couplings are positive and one is negative. An opposite situation, in which two couplings are negative and only one is positive, is observed in Domains \circled{3} and \circled{5}, therefore giving rise to a stripy antiferromagnetic state. It is worth noting that these magnetic phases can be realized within a relatively narrow interval of lattice strain of $\sim$5\% in magnitude -- provided that the direction along which the deformation occurs is carefully controlled --  thereby highlighting a marked interplay between lattice deformations and magnetic exchange couplings in monolayer CrI$_3$.  Indeed, the six strain-induced magnetic transitions encompass all the possible magnetically ordered phases which can be realized on a honeycomb lattice.

Importantly, we remark that the phase diagram shown in Fig.\ \ref{Fig3} displays a singularity located at $\epsilon_a \approx \epsilon_b \approx -5 \%$. This corresponds to the point in which the domain boundaries intersect, leading to   $J_1 \approx J_2 \approx J_3 \approx 0$.  According to the adopted isotropic Heisenberg spin Hamiltonian, such a critical point signals the absence of any magnetic order.  The strain-induced quenching of the isotropic exchange couplings may be exploited to single out the otherwise negligible anisotropic exchange couplings in monolayer CrI$_3$  \cite{Pizzochero19}. This is of particular interest in the context of Kitaev physics \cite{Kitaev}, which emerges when a large Kitaev-to-Heisenberg coupling ratio is achieved. In our previous work \cite{Pizzochero19}, we have quantified the Kitaev parameter $K$ in unstrained monolayer CrI$_3$ to be $-0.08$ meV, resulting in a practically negligible $K/J$ ratio of 0.06.  Under appropriate strain conditions ($\epsilon_a \approx \epsilon_b \approx -5 \%$), however, we suggest that the $K/J$ ratio can be enhanced to a great extent, hence driving monolayer CrI$_3$ in the Kitaev spin-liquid phase, provided that $K$ remains non-vanishing. Indeed, a recent quantum chemistry investigation has demonstrated that compressive biaxial strain substantially increases the magnitude of the ferromagnetic Kitaev coupling in several honeycomb lattices \cite{Yadav18b}.

 \begin{figure}[]
  \centering
  \includegraphics[width=0.9\columnwidth]{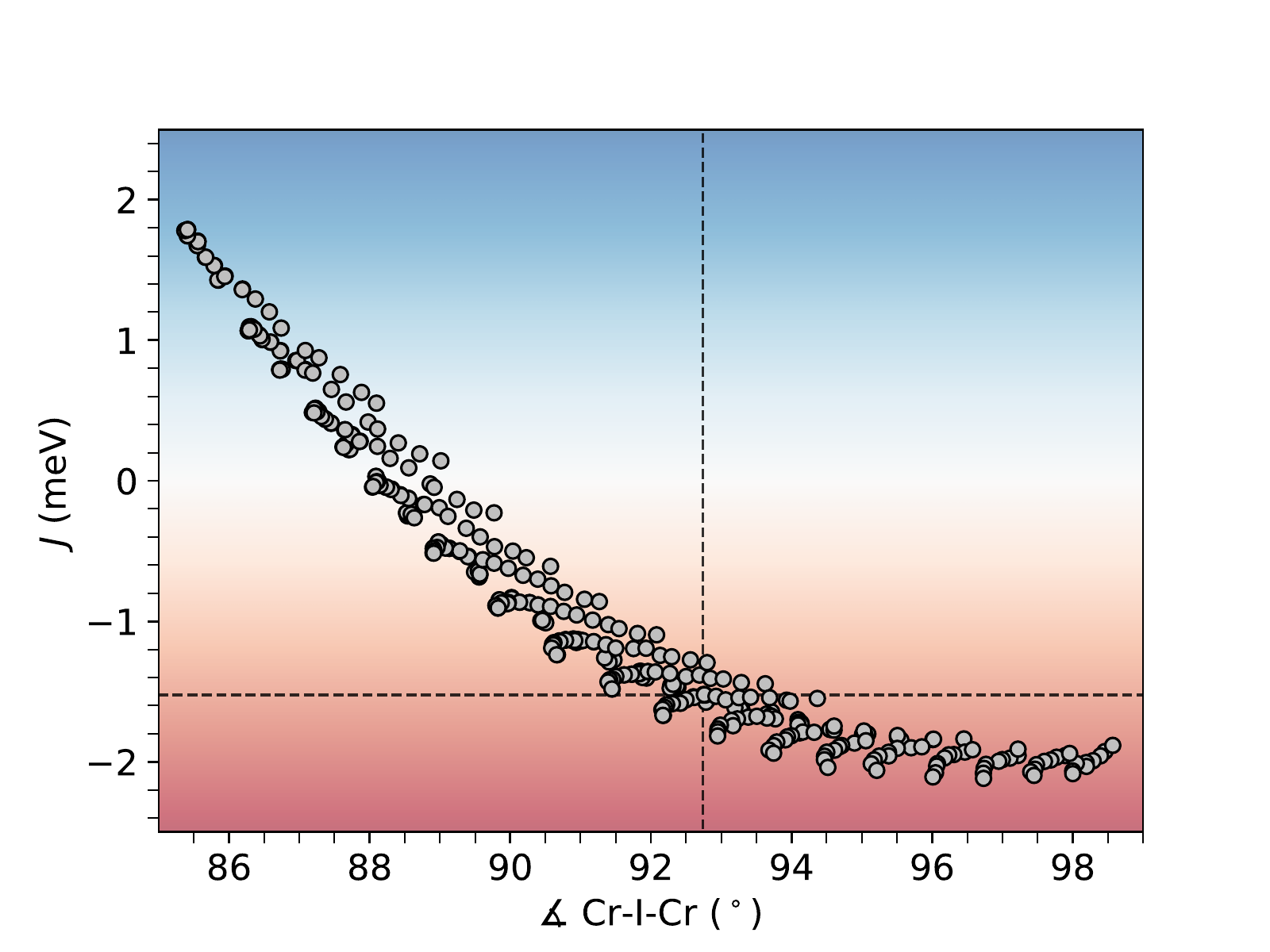}
  \caption{Exchange couplings ($J$) as a function of the metal-ligand-metal bond angle ($\measuredangle$Cr-I-Cr) in monolayer CrI$_3$. Dashed lines mark the equilibrium values in unstrained CrI$_3$. Background shaded in red and blue colors marks positive and negative values of $J$, respectively,  indicating a ferromagnetic and antiferromagnetic interaction between nearest-neighbors Cr$^{3+}$ ions, respectively.}
  \label{Fig4}
\end{figure}

Finally, we briefly comment on the dependence of the exchange couplings on the crystal structure. According to the rule proposed by Goodenough \cite{Goodenough58} and Kanamori \cite{Kanamori59}, the sign of the exchange coupling in superexchange-dominated magnetic lattices is mainly governed by the metal-ligand-metal angle. Specifically, depending on whether such an angle is close to either $90^{\circ}$ or $180^{\circ}$, a parallel or antiparallel spin orientation has to be expected, respectively. Unstrained CrI$_3$ complies with this rule \cite{Lado17}, as the bond angle $\measuredangle$Cr-I-Cr $\approx 90^{\circ}$ is accompanied by $J < 0$. In Fig.\ \ref{Fig4}, we present the dependence of the 459 isotropic exchange couplings calculated in this work on the $\measuredangle$Cr-I-Cr. We observe that a ferro- to antiferro-magnetic crossover occurs when the bond angle takes values which are lower than $88^{\circ}$. This is at odds with the Goodenough-Kanamori rule, according to which such a crossover should instead take place when approaching the $180^{\circ}$ limit.  Overall, our results indicate that the Goodenough-Kanamori rule is not an appropriate guideline to rationalize the magnetism of monolayer CrI$_3$ subjected to lattice deformations.

\medskip
\paragraph{Conclusion.} 
In summary, we have combined model spin Hamiltonians with DFT+$U$ calculations to address the magnetic response of monolayer CrI$_3$  to in-plane lattice deformations. We revealed that, depending on the magnitude and the direction along which the applied strain is exerted, the ferromagnetic ground state can be driven into either the N\'eel antiferromagnetic or ferrimagnetic phases, these latter including both stripy and zigzag spin configurations. We identified a critical point emerging at the intersection between the magnetic phase boundaries in the strain space whereby the isotropic Heisenberg exchange couplings vanish, hence quenching any magnetically ordered phase. To conclude, our findings open new viable routes for extensively engineering the magnetism of CrI$_3$ through lattice strain down to the ultimate limit of thickness miniaturization.


\medskip
\paragraph{Acknowledgments.} 
M.P.\ acknowledges Ravi Yadav for fruitful interactions. This work was financially supported by the Swiss National Science Foundation (Grant No.\ 172543). First-principles calculations were performed at the Swiss National Supercomputing Center (CSCS) under the project s832.

 \bibliography{References}

\begin{thebibliography}{44}
\expandafter\ifx\csname natexlab\endcsname\relax\def\natexlab#1{#1}\fi
\expandafter\ifx\csname bibnamefont\endcsname\relax
  \def\bibnamefont#1{#1}\fi
\expandafter\ifx\csname bibfnamefont\endcsname\relax
  \def\bibfnamefont#1{#1}\fi
\expandafter\ifx\csname citenamefont\endcsname\relax
  \def\citenamefont#1{#1}\fi
\expandafter\ifx\csname url\endcsname\relax
  \def\url#1{\texttt{#1}}\fi
\expandafter\ifx\csname urlprefix\endcsname\relax\def\urlprefix{URL }\fi
\providecommand{\bibinfo}[2]{#2}
\providecommand{\eprint}[2][]{\url{#2}}

\bibitem[{\citenamefont{Novoselov et~al.}(2004)\citenamefont{Novoselov, Geim,
  Morozov, Jiang, Zhang, Dubonos, Grigorieva, and Firsov}}]{Novoselov04}
\bibinfo{author}{\bibfnamefont{K.~S.} \bibnamefont{Novoselov}},
  \bibinfo{author}{\bibfnamefont{A.~K.} \bibnamefont{Geim}},
  \bibinfo{author}{\bibfnamefont{S.~V.} \bibnamefont{Morozov}},
  \bibinfo{author}{\bibfnamefont{D.}~\bibnamefont{Jiang}},
  \bibinfo{author}{\bibfnamefont{Y.}~\bibnamefont{Zhang}},
  \bibinfo{author}{\bibfnamefont{S.~V.} \bibnamefont{Dubonos}},
  \bibinfo{author}{\bibfnamefont{I.~V.} \bibnamefont{Grigorieva}},
  \bibnamefont{and} \bibinfo{author}{\bibfnamefont{A.~A.}
  \bibnamefont{Firsov}}, \bibinfo{journal}{Science}
  \textbf{\bibinfo{volume}{306}}, \bibinfo{pages}{666} (\bibinfo{year}{2004}).

\bibitem[{\citenamefont{Novoselov et~al.}(2005)\citenamefont{Novoselov, Jiang,
  Schedin, Booth, Khotkevich, Morozov, and Geim}}]{Novoselov05}
\bibinfo{author}{\bibfnamefont{K.~S.} \bibnamefont{Novoselov}},
  \bibinfo{author}{\bibfnamefont{D.}~\bibnamefont{Jiang}},
  \bibinfo{author}{\bibfnamefont{F.}~\bibnamefont{Schedin}},
  \bibinfo{author}{\bibfnamefont{T.~J.} \bibnamefont{Booth}},
  \bibinfo{author}{\bibfnamefont{V.~V.} \bibnamefont{Khotkevich}},
  \bibinfo{author}{\bibfnamefont{S.~V.} \bibnamefont{Morozov}},
  \bibnamefont{and} \bibinfo{author}{\bibfnamefont{A.~K.} \bibnamefont{Geim}},
  \bibinfo{journal}{Proceedings of the National Academy of Sciences of the
  United States of America} \textbf{\bibinfo{volume}{102}},
  \bibinfo{pages}{10451} (\bibinfo{year}{2005}).

\bibitem[{\citenamefont{Nicolosi et~al.}(2013)\citenamefont{Nicolosi,
  Chhowalla, Kanatzidis, Strano, and Coleman}}]{Nicolosi13}
\bibinfo{author}{\bibfnamefont{V.}~\bibnamefont{Nicolosi}},
  \bibinfo{author}{\bibfnamefont{M.}~\bibnamefont{Chhowalla}},
  \bibinfo{author}{\bibfnamefont{M.~G.} \bibnamefont{Kanatzidis}},
  \bibinfo{author}{\bibfnamefont{M.~S.} \bibnamefont{Strano}},
  \bibnamefont{and} \bibinfo{author}{\bibfnamefont{J.~N.}
  \bibnamefont{Coleman}}, \bibinfo{journal}{Science}
  \textbf{\bibinfo{volume}{340}}, \bibinfo{pages}{1226419}
  (\bibinfo{year}{2013}).

\bibitem[{\citenamefont{Mounet et~al.}(2018)\citenamefont{Mounet, Gibertini,
  Schwaller, Campi, Merkys, Marrazzo, Sohier, Castelli, Cepellotti, Pizzi
  et~al.}}]{Mounet18}
\bibinfo{author}{\bibfnamefont{N.}~\bibnamefont{Mounet}},
  \bibinfo{author}{\bibfnamefont{M.}~\bibnamefont{Gibertini}},
  \bibinfo{author}{\bibfnamefont{P.}~\bibnamefont{Schwaller}},
  \bibinfo{author}{\bibfnamefont{D.}~\bibnamefont{Campi}},
  \bibinfo{author}{\bibfnamefont{A.}~\bibnamefont{Merkys}},
  \bibinfo{author}{\bibfnamefont{A.}~\bibnamefont{Marrazzo}},
  \bibinfo{author}{\bibfnamefont{T.}~\bibnamefont{Sohier}},
  \bibinfo{author}{\bibfnamefont{I.~E.} \bibnamefont{Castelli}},
  \bibinfo{author}{\bibfnamefont{A.}~\bibnamefont{Cepellotti}},
  \bibinfo{author}{\bibfnamefont{G.}~\bibnamefont{Pizzi}},
  \bibnamefont{et~al.}, \bibinfo{journal}{Nature Nanotechnology}
  \textbf{\bibinfo{volume}{13}}, \bibinfo{pages}{246} (\bibinfo{year}{2018}).

\bibitem[{\citenamefont{Castro~Neto et~al.}(2009)\citenamefont{Castro~Neto,
  Guinea, Peres, Novoselov, and Geim}}]{CastroNeto09}
\bibinfo{author}{\bibfnamefont{A.~H.} \bibnamefont{Castro~Neto}},
  \bibinfo{author}{\bibfnamefont{F.}~\bibnamefont{Guinea}},
  \bibinfo{author}{\bibfnamefont{N.~M.~R.} \bibnamefont{Peres}},
  \bibinfo{author}{\bibfnamefont{K.~S.} \bibnamefont{Novoselov}},
  \bibnamefont{and} \bibinfo{author}{\bibfnamefont{A.~K.} \bibnamefont{Geim}},
  \bibinfo{journal}{Review of Modern Physics} \textbf{\bibinfo{volume}{81}},
  \bibinfo{pages}{109} (\bibinfo{year}{2009}).

\bibitem[{\citenamefont{Yazyev and Kis}(2015)}]{Yazyev15}
\bibinfo{author}{\bibfnamefont{O.~V.} \bibnamefont{Yazyev}} \bibnamefont{and}
  \bibinfo{author}{\bibfnamefont{A.}~\bibnamefont{Kis}},
  \bibinfo{journal}{Materials Today} \textbf{\bibinfo{volume}{18}},
  \bibinfo{pages}{20} (\bibinfo{year}{2015}).

\bibitem[{\citenamefont{Chen et~al.}(2018)\citenamefont{Chen, Kim, Bernard,
  Pizzochero, Zaldivar, Pascual, Ugeda, Yazyev, Greber, Osterwalder
  et~al.}}]{Chen2018}
\bibinfo{author}{\bibfnamefont{M.-W.} \bibnamefont{Chen}},
  \bibinfo{author}{\bibfnamefont{H.}~\bibnamefont{Kim}},
  \bibinfo{author}{\bibfnamefont{C.}~\bibnamefont{Bernard}},
  \bibinfo{author}{\bibfnamefont{M.}~\bibnamefont{Pizzochero}},
  \bibinfo{author}{\bibfnamefont{J.}~\bibnamefont{Zaldivar}},
  \bibinfo{author}{\bibfnamefont{J.~I.} \bibnamefont{Pascual}},
  \bibinfo{author}{\bibfnamefont{M.~M.} \bibnamefont{Ugeda}},
  \bibinfo{author}{\bibfnamefont{O.~V.} \bibnamefont{Yazyev}},
  \bibinfo{author}{\bibfnamefont{T.}~\bibnamefont{Greber}},
  \bibinfo{author}{\bibfnamefont{J.}~\bibnamefont{Osterwalder}},
  \bibnamefont{et~al.}, \bibinfo{journal}{ACS Nano}
  \textbf{\bibinfo{volume}{12}}, \bibinfo{pages}{11161} (\bibinfo{year}{2018}).

\bibitem[{\citenamefont{Pedramrazi et~al.}(2019)\citenamefont{Pedramrazi,
  Herbig, Pulkin, Tang, Phillips, Wong, Ryu, Pizzochero, Chen, Wang
  et~al.}}]{Pedramrazi19}
\bibinfo{author}{\bibfnamefont{Z.}~\bibnamefont{Pedramrazi}},
  \bibinfo{author}{\bibfnamefont{C.}~\bibnamefont{Herbig}},
  \bibinfo{author}{\bibfnamefont{A.}~\bibnamefont{Pulkin}},
  \bibinfo{author}{\bibfnamefont{S.}~\bibnamefont{Tang}},
  \bibinfo{author}{\bibfnamefont{M.}~\bibnamefont{Phillips}},
  \bibinfo{author}{\bibfnamefont{D.}~\bibnamefont{Wong}},
  \bibinfo{author}{\bibfnamefont{H.}~\bibnamefont{Ryu}},
  \bibinfo{author}{\bibfnamefont{M.}~\bibnamefont{Pizzochero}},
  \bibinfo{author}{\bibfnamefont{Y.}~\bibnamefont{Chen}},
  \bibinfo{author}{\bibfnamefont{F.}~\bibnamefont{Wang}}, \bibnamefont{et~al.},
  \bibinfo{journal}{Nano Letters} \textbf{\bibinfo{volume}{19}},
  \bibinfo{pages}{5634} (\bibinfo{year}{2019}).

\bibitem[{\citenamefont{Castro~Neto}(2001)}]{CastroNeto01}
\bibinfo{author}{\bibfnamefont{A.~H.} \bibnamefont{Castro~Neto}},
  \bibinfo{journal}{Physical Review Letters} \textbf{\bibinfo{volume}{86}},
  \bibinfo{pages}{4382} (\bibinfo{year}{2001}).

\bibitem[{\citenamefont{Cao et~al.}(2018{\natexlab{a}})\citenamefont{Cao,
  Fatemi, Fang, Watanabe, Taniguchi, Kaxiras, and Jarillo-Herrero}}]{Cao18a}
\bibinfo{author}{\bibfnamefont{Y.}~\bibnamefont{Cao}},
  \bibinfo{author}{\bibfnamefont{V.}~\bibnamefont{Fatemi}},
  \bibinfo{author}{\bibfnamefont{S.}~\bibnamefont{Fang}},
  \bibinfo{author}{\bibfnamefont{K.}~\bibnamefont{Watanabe}},
  \bibinfo{author}{\bibfnamefont{T.}~\bibnamefont{Taniguchi}},
  \bibinfo{author}{\bibfnamefont{E.}~\bibnamefont{Kaxiras}}, \bibnamefont{and}
  \bibinfo{author}{\bibfnamefont{P.}~\bibnamefont{Jarillo-Herrero}},
  \bibinfo{journal}{Nature} \textbf{\bibinfo{volume}{556}}, \bibinfo{pages}{43}
  (\bibinfo{year}{2018}{\natexlab{a}}).

\bibitem[{\citenamefont{Cao et~al.}(2018{\natexlab{b}})\citenamefont{Cao,
  Fatemi, Demir, Fang, Tomarken, Luo, Sanchez-Yamagishi, Watanabe, Taniguchi,
  Kaxiras et~al.}}]{Cao18b}
\bibinfo{author}{\bibfnamefont{Y.}~\bibnamefont{Cao}},
  \bibinfo{author}{\bibfnamefont{V.}~\bibnamefont{Fatemi}},
  \bibinfo{author}{\bibfnamefont{A.}~\bibnamefont{Demir}},
  \bibinfo{author}{\bibfnamefont{S.}~\bibnamefont{Fang}},
  \bibinfo{author}{\bibfnamefont{S.~L.} \bibnamefont{Tomarken}},
  \bibinfo{author}{\bibfnamefont{J.~Y.} \bibnamefont{Luo}},
  \bibinfo{author}{\bibfnamefont{J.~D.} \bibnamefont{Sanchez-Yamagishi}},
  \bibinfo{author}{\bibfnamefont{K.}~\bibnamefont{Watanabe}},
  \bibinfo{author}{\bibfnamefont{T.}~\bibnamefont{Taniguchi}},
  \bibinfo{author}{\bibfnamefont{E.}~\bibnamefont{Kaxiras}},
  \bibnamefont{et~al.}, \bibinfo{journal}{Nature}
  \textbf{\bibinfo{volume}{556}}, \bibinfo{pages}{80}
  (\bibinfo{year}{2018}{\natexlab{b}}).

\bibitem[{\citenamefont{Manzeli et~al.}(2017)\citenamefont{Manzeli,
  Ovchinnikov, Pasquier, Yazyev, and Kis}}]{Manzeli17}
\bibinfo{author}{\bibfnamefont{S.}~\bibnamefont{Manzeli}},
  \bibinfo{author}{\bibfnamefont{D.}~\bibnamefont{Ovchinnikov}},
  \bibinfo{author}{\bibfnamefont{D.}~\bibnamefont{Pasquier}},
  \bibinfo{author}{\bibfnamefont{O.~V.} \bibnamefont{Yazyev}},
  \bibnamefont{and} \bibinfo{author}{\bibfnamefont{A.}~\bibnamefont{Kis}},
  \bibinfo{journal}{Nature Reviews Materials} \textbf{\bibinfo{volume}{2}},
  \bibinfo{pages}{17033} (\bibinfo{year}{2017}).

\bibitem[{\citenamefont{Pizzochero
  et~al.}(2019{\natexlab{a}})\citenamefont{Pizzochero, Bonfanti, and
  Martinazzo}}]{Pizzochero19bis}
\bibinfo{author}{\bibfnamefont{M.}~\bibnamefont{Pizzochero}},
  \bibinfo{author}{\bibfnamefont{M.}~\bibnamefont{Bonfanti}}, \bibnamefont{and}
  \bibinfo{author}{\bibfnamefont{R.}~\bibnamefont{Martinazzo}},
  \bibinfo{journal}{Physical Chemistry Chemical Physics}
  \textbf{\bibinfo{volume}{21}}, \bibinfo{pages}{26342}
  (\bibinfo{year}{2019}{\natexlab{a}}).

\bibitem[{\citenamefont{Burch et~al.}(2018)\citenamefont{Burch, Mandrus, and
  Park}}]{Burch2018}
\bibinfo{author}{\bibfnamefont{K.~S.} \bibnamefont{Burch}},
  \bibinfo{author}{\bibfnamefont{D.}~\bibnamefont{Mandrus}}, \bibnamefont{and}
  \bibinfo{author}{\bibfnamefont{J.-G.} \bibnamefont{Park}},
  \bibinfo{journal}{Nature} \textbf{\bibinfo{volume}{563}}, \bibinfo{pages}{47}
  (\bibinfo{year}{2018}).

\bibitem[{\citenamefont{Gong and Zhang}(2019)}]{Gong19}
\bibinfo{author}{\bibfnamefont{C.}~\bibnamefont{Gong}} \bibnamefont{and}
  \bibinfo{author}{\bibfnamefont{X.}~\bibnamefont{Zhang}},
  \bibinfo{journal}{Science} \textbf{\bibinfo{volume}{363}},
  \bibinfo{pages}{706} (\bibinfo{year}{2019}).

\bibitem[{\citenamefont{Mermin and Wagner}(1966)}]{Mermin66}
\bibinfo{author}{\bibfnamefont{N.~D.} \bibnamefont{Mermin}} \bibnamefont{and}
  \bibinfo{author}{\bibfnamefont{H.}~\bibnamefont{Wagner}},
  \bibinfo{journal}{Physical Review Letters} \textbf{\bibinfo{volume}{17}},
  \bibinfo{pages}{1133} (\bibinfo{year}{1966}).

\bibitem[{\citenamefont{Huang et~al.}(2017)\citenamefont{Huang, Clark,
  Navarro-Moratalla, Klein, Cheng, Seyler, Zhong, Schmidgall, McGuire, Cobden
  et~al.}}]{Huang2017}
\bibinfo{author}{\bibfnamefont{B.}~\bibnamefont{Huang}},
  \bibinfo{author}{\bibfnamefont{G.}~\bibnamefont{Clark}},
  \bibinfo{author}{\bibfnamefont{E.}~\bibnamefont{Navarro-Moratalla}},
  \bibinfo{author}{\bibfnamefont{D.~R.} \bibnamefont{Klein}},
  \bibinfo{author}{\bibfnamefont{R.}~\bibnamefont{Cheng}},
  \bibinfo{author}{\bibfnamefont{K.~L.} \bibnamefont{Seyler}},
  \bibinfo{author}{\bibfnamefont{D.}~\bibnamefont{Zhong}},
  \bibinfo{author}{\bibfnamefont{E.}~\bibnamefont{Schmidgall}},
  \bibinfo{author}{\bibfnamefont{M.~A.} \bibnamefont{McGuire}},
  \bibinfo{author}{\bibfnamefont{D.~H.} \bibnamefont{Cobden}},
  \bibnamefont{et~al.}, \bibinfo{journal}{Nature}
  \textbf{\bibinfo{volume}{546}}, \bibinfo{pages}{270} (\bibinfo{year}{2017}).

\bibitem[{\citenamefont{Gibertini et~al.}(2019)\citenamefont{Gibertini,
  Koperski, Morpurgo, and Novoselov}}]{Gibertini19}
\bibinfo{author}{\bibfnamefont{M.}~\bibnamefont{Gibertini}},
  \bibinfo{author}{\bibfnamefont{M.}~\bibnamefont{Koperski}},
  \bibinfo{author}{\bibfnamefont{A.~F.} \bibnamefont{Morpurgo}},
  \bibnamefont{and} \bibinfo{author}{\bibfnamefont{K.~S.}
  \bibnamefont{Novoselov}}, \bibinfo{journal}{Nature Nanotechnology}
  \textbf{\bibinfo{volume}{14}}, \bibinfo{pages}{408} (\bibinfo{year}{2019}).

\bibitem[{\citenamefont{Torelli et~al.}(2019)\citenamefont{Torelli, Thygesen,
  and Olsen}}]{Torelli2019}
\bibinfo{author}{\bibfnamefont{D.}~\bibnamefont{Torelli}},
  \bibinfo{author}{\bibfnamefont{K.~S.} \bibnamefont{Thygesen}},
  \bibnamefont{and} \bibinfo{author}{\bibfnamefont{T.}~\bibnamefont{Olsen}},
  \bibinfo{journal}{2D Materials} \textbf{\bibinfo{volume}{6}},
  \bibinfo{pages}{045018} (\bibinfo{year}{2019}).

\bibitem[{\citenamefont{Avsar et~al.}(2019)\citenamefont{Avsar, Ciarrocchi,
  Pizzochero, Unuchek, Yazyev, and Kis}}]{Avsar19}
\bibinfo{author}{\bibfnamefont{A.}~\bibnamefont{Avsar}},
  \bibinfo{author}{\bibfnamefont{A.}~\bibnamefont{Ciarrocchi}},
  \bibinfo{author}{\bibfnamefont{M.}~\bibnamefont{Pizzochero}},
  \bibinfo{author}{\bibfnamefont{D.}~\bibnamefont{Unuchek}},
  \bibinfo{author}{\bibfnamefont{O.~V.} \bibnamefont{Yazyev}},
  \bibnamefont{and} \bibinfo{author}{\bibfnamefont{A.}~\bibnamefont{Kis}},
  \bibinfo{journal}{Nature Nanotechnology} \textbf{\bibinfo{volume}{14}},
  \bibinfo{pages}{674} (\bibinfo{year}{2019}).

\bibitem[{\citenamefont{McGuire et~al.}(2015)\citenamefont{McGuire, Dixit,
  Cooper, and Sales}}]{McGuire15}
\bibinfo{author}{\bibfnamefont{M.~A.} \bibnamefont{McGuire}},
  \bibinfo{author}{\bibfnamefont{H.}~\bibnamefont{Dixit}},
  \bibinfo{author}{\bibfnamefont{V.~R.} \bibnamefont{Cooper}},
  \bibnamefont{and} \bibinfo{author}{\bibfnamefont{B.~C.} \bibnamefont{Sales}},
  \bibinfo{journal}{Chemistry of Materials} \textbf{\bibinfo{volume}{27}},
  \bibinfo{pages}{612} (\bibinfo{year}{2015}).

\bibitem[{\citenamefont{Lado and Fern{\'{a}}ndez-Rossier}(2017)}]{Lado17}
\bibinfo{author}{\bibfnamefont{J.~L.} \bibnamefont{Lado}} \bibnamefont{and}
  \bibinfo{author}{\bibfnamefont{J.}~\bibnamefont{Fern{\'{a}}ndez-Rossier}},
  \bibinfo{journal}{2D Materials} \textbf{\bibinfo{volume}{4}},
  \bibinfo{pages}{035002} (\bibinfo{year}{2017}).

\bibitem[{\citenamefont{Huang et~al.}(2018)\citenamefont{Huang, Clark, Klein,
  MacNeill, Navarro-Moratalla, Seyler, Wilson, McGuire, Cobden, Xiao
  et~al.}}]{Huang2018}
\bibinfo{author}{\bibfnamefont{B.}~\bibnamefont{Huang}},
  \bibinfo{author}{\bibfnamefont{G.}~\bibnamefont{Clark}},
  \bibinfo{author}{\bibfnamefont{D.~R.} \bibnamefont{Klein}},
  \bibinfo{author}{\bibfnamefont{D.}~\bibnamefont{MacNeill}},
  \bibinfo{author}{\bibfnamefont{E.}~\bibnamefont{Navarro-Moratalla}},
  \bibinfo{author}{\bibfnamefont{K.~L.} \bibnamefont{Seyler}},
  \bibinfo{author}{\bibfnamefont{N.}~\bibnamefont{Wilson}},
  \bibinfo{author}{\bibfnamefont{M.~A.} \bibnamefont{McGuire}},
  \bibinfo{author}{\bibfnamefont{D.~H.} \bibnamefont{Cobden}},
  \bibinfo{author}{\bibfnamefont{D.}~\bibnamefont{Xiao}}, \bibnamefont{et~al.},
  \bibinfo{journal}{Nature Nanotechnology} \textbf{\bibinfo{volume}{13}},
  \bibinfo{pages}{544} (\bibinfo{year}{2018}).

\bibitem[{\citenamefont{Jiang et~al.}(2018)\citenamefont{Jiang, Li, Wang, Mak,
  and Shan}}]{Jiang2018}
\bibinfo{author}{\bibfnamefont{S.}~\bibnamefont{Jiang}},
  \bibinfo{author}{\bibfnamefont{L.}~\bibnamefont{Li}},
  \bibinfo{author}{\bibfnamefont{Z.}~\bibnamefont{Wang}},
  \bibinfo{author}{\bibfnamefont{K.~F.} \bibnamefont{Mak}}, \bibnamefont{and}
  \bibinfo{author}{\bibfnamefont{J.}~\bibnamefont{Shan}},
  \bibinfo{journal}{Nature Nanotechnology} \textbf{\bibinfo{volume}{13}},
  \bibinfo{pages}{549} (\bibinfo{year}{2018}).

\bibitem[{\citenamefont{Song et~al.}(2019)\citenamefont{Song, Fei, Yankowitz,
  Lin, Jiang, Hwangbo, Zhang, Sun, Taniguchi, Watanabe et~al.}}]{Song19}
\bibinfo{author}{\bibfnamefont{T.}~\bibnamefont{Song}},
  \bibinfo{author}{\bibfnamefont{Z.}~\bibnamefont{Fei}},
  \bibinfo{author}{\bibfnamefont{M.}~\bibnamefont{Yankowitz}},
  \bibinfo{author}{\bibfnamefont{Z.}~\bibnamefont{Lin}},
  \bibinfo{author}{\bibfnamefont{Q.}~\bibnamefont{Jiang}},
  \bibinfo{author}{\bibfnamefont{K.}~\bibnamefont{Hwangbo}},
  \bibinfo{author}{\bibfnamefont{Q.}~\bibnamefont{Zhang}},
  \bibinfo{author}{\bibfnamefont{B.}~\bibnamefont{Sun}},
  \bibinfo{author}{\bibfnamefont{T.}~\bibnamefont{Taniguchi}},
  \bibinfo{author}{\bibfnamefont{K.}~\bibnamefont{Watanabe}},
  \bibnamefont{et~al.}, \bibinfo{journal}{arXiv:1905.10860}
  (\bibinfo{year}{2019}).

\bibitem[{\citenamefont{Li et~al.}(2019)\citenamefont{Li, Jiang, Sivadas, Wang,
  Xu, Weber, Goldberger, Watanabe, Taniguchi, Fennie et~al.}}]{Tingxin19}
\bibinfo{author}{\bibfnamefont{T.}~\bibnamefont{Li}},
  \bibinfo{author}{\bibfnamefont{S.}~\bibnamefont{Jiang}},
  \bibinfo{author}{\bibfnamefont{N.}~\bibnamefont{Sivadas}},
  \bibinfo{author}{\bibfnamefont{Z.}~\bibnamefont{Wang}},
  \bibinfo{author}{\bibfnamefont{Y.}~\bibnamefont{Xu}},
  \bibinfo{author}{\bibfnamefont{D.}~\bibnamefont{Weber}},
  \bibinfo{author}{\bibfnamefont{J.~E.} \bibnamefont{Goldberger}},
  \bibinfo{author}{\bibfnamefont{K.}~\bibnamefont{Watanabe}},
  \bibinfo{author}{\bibfnamefont{T.}~\bibnamefont{Taniguchi}},
  \bibinfo{author}{\bibfnamefont{C.~J.} \bibnamefont{Fennie}},
  \bibnamefont{et~al.}, \bibinfo{journal}{arXiv:1905.10905}
  (\bibinfo{year}{2019}).

\bibitem[{\citenamefont{Thiel et~al.}(2019)\citenamefont{Thiel, Wang, Tschudin,
  Rohner, Guti{\'e}rrez-Lezama, Ubrig, Gibertini, Giannini, Morpurgo, and
  Maletinsky}}]{Thiel19}
\bibinfo{author}{\bibfnamefont{L.}~\bibnamefont{Thiel}},
  \bibinfo{author}{\bibfnamefont{Z.}~\bibnamefont{Wang}},
  \bibinfo{author}{\bibfnamefont{M.~A.} \bibnamefont{Tschudin}},
  \bibinfo{author}{\bibfnamefont{D.}~\bibnamefont{Rohner}},
  \bibinfo{author}{\bibfnamefont{I.}~\bibnamefont{Guti{\'e}rrez-Lezama}},
  \bibinfo{author}{\bibfnamefont{N.}~\bibnamefont{Ubrig}},
  \bibinfo{author}{\bibfnamefont{M.}~\bibnamefont{Gibertini}},
  \bibinfo{author}{\bibfnamefont{E.}~\bibnamefont{Giannini}},
  \bibinfo{author}{\bibfnamefont{A.~F.} \bibnamefont{Morpurgo}},
  \bibnamefont{and}
  \bibinfo{author}{\bibfnamefont{P.}~\bibnamefont{Maletinsky}},
  \bibinfo{journal}{Science} \textbf{\bibinfo{volume}{364}},
  \bibinfo{pages}{973} (\bibinfo{year}{2019}).

\bibitem[{\citenamefont{Sivadas et~al.}(2018)\citenamefont{Sivadas, Okamoto,
  Xu, Fennie, and Xiao}}]{Sivadas18}
\bibinfo{author}{\bibfnamefont{N.}~\bibnamefont{Sivadas}},
  \bibinfo{author}{\bibfnamefont{S.}~\bibnamefont{Okamoto}},
  \bibinfo{author}{\bibfnamefont{X.}~\bibnamefont{Xu}},
  \bibinfo{author}{\bibfnamefont{C.~J.} \bibnamefont{Fennie}},
  \bibnamefont{and} \bibinfo{author}{\bibfnamefont{D.}~\bibnamefont{Xiao}},
  \bibinfo{journal}{Nano Letters} \textbf{\bibinfo{volume}{18}},
  \bibinfo{pages}{7658} (\bibinfo{year}{2018}).

\bibitem[{\citenamefont{Ubrig et~al.}(2019)\citenamefont{Ubrig, Wang, Teyssier,
  Taniguchi, Watanabe, Giannini, Morpurgo, and Gibertini}}]{Ubrig19}
\bibinfo{author}{\bibfnamefont{N.}~\bibnamefont{Ubrig}},
  \bibinfo{author}{\bibfnamefont{Z.}~\bibnamefont{Wang}},
  \bibinfo{author}{\bibfnamefont{J.}~\bibnamefont{Teyssier}},
  \bibinfo{author}{\bibfnamefont{T.}~\bibnamefont{Taniguchi}},
  \bibinfo{author}{\bibfnamefont{K.}~\bibnamefont{Watanabe}},
  \bibinfo{author}{\bibfnamefont{E.}~\bibnamefont{Giannini}},
  \bibinfo{author}{\bibfnamefont{A.~F.} \bibnamefont{Morpurgo}},
  \bibnamefont{and}
  \bibinfo{author}{\bibfnamefont{M.}~\bibnamefont{Gibertini}},
  \bibinfo{journal}{2D Materials} \textbf{\bibinfo{volume}{7}},
  \bibinfo{pages}{015007} (\bibinfo{year}{2019}).

\bibitem[{\citenamefont{Wang et~al.}(2018)\citenamefont{Wang,
  Guti{\'e}rrez-Lezama, Ubrig, Kroner, Gibertini, Taniguchi, Watanabe,
  Imamoglu, Giannini, and Morpurgo}}]{Wang2018}
\bibinfo{author}{\bibfnamefont{Z.}~\bibnamefont{Wang}},
  \bibinfo{author}{\bibfnamefont{I.}~\bibnamefont{Guti{\'e}rrez-Lezama}},
  \bibinfo{author}{\bibfnamefont{N.}~\bibnamefont{Ubrig}},
  \bibinfo{author}{\bibfnamefont{M.}~\bibnamefont{Kroner}},
  \bibinfo{author}{\bibfnamefont{M.}~\bibnamefont{Gibertini}},
  \bibinfo{author}{\bibfnamefont{T.}~\bibnamefont{Taniguchi}},
  \bibinfo{author}{\bibfnamefont{K.}~\bibnamefont{Watanabe}},
  \bibinfo{author}{\bibfnamefont{A.}~\bibnamefont{Imamoglu}},
  \bibinfo{author}{\bibfnamefont{E.}~\bibnamefont{Giannini}}, \bibnamefont{and}
  \bibinfo{author}{\bibfnamefont{A.~F.} \bibnamefont{Morpurgo}},
  \bibinfo{journal}{Nature Communications} \textbf{\bibinfo{volume}{9}},
  \bibinfo{pages}{2516} (\bibinfo{year}{2018}).

\bibitem[{\citenamefont{Song et~al.}(2018)\citenamefont{Song, Cai, Tu, Zhang,
  Huang, Wilson, Seyler, Zhu, Taniguchi, Watanabe et~al.}}]{Song1214}
\bibinfo{author}{\bibfnamefont{T.}~\bibnamefont{Song}},
  \bibinfo{author}{\bibfnamefont{X.}~\bibnamefont{Cai}},
  \bibinfo{author}{\bibfnamefont{M.~W.-Y.} \bibnamefont{Tu}},
  \bibinfo{author}{\bibfnamefont{X.}~\bibnamefont{Zhang}},
  \bibinfo{author}{\bibfnamefont{B.}~\bibnamefont{Huang}},
  \bibinfo{author}{\bibfnamefont{N.~P.} \bibnamefont{Wilson}},
  \bibinfo{author}{\bibfnamefont{K.~L.} \bibnamefont{Seyler}},
  \bibinfo{author}{\bibfnamefont{L.}~\bibnamefont{Zhu}},
  \bibinfo{author}{\bibfnamefont{T.}~\bibnamefont{Taniguchi}},
  \bibinfo{author}{\bibfnamefont{K.}~\bibnamefont{Watanabe}},
  \bibnamefont{et~al.}, \bibinfo{journal}{Science}
  \textbf{\bibinfo{volume}{360}}, \bibinfo{pages}{1214} (\bibinfo{year}{2018}).

\bibitem[{\citenamefont{Klein et~al.}(2018)\citenamefont{Klein, MacNeill, Lado,
  Soriano, Navarro-Moratalla, Watanabe, Taniguchi, Manni, Canfield,
  Fern{\'a}ndez~Rossier et~al.}}]{Klein1218}
\bibinfo{author}{\bibfnamefont{D.~R.} \bibnamefont{Klein}},
  \bibinfo{author}{\bibfnamefont{D.}~\bibnamefont{MacNeill}},
  \bibinfo{author}{\bibfnamefont{J.~L.} \bibnamefont{Lado}},
  \bibinfo{author}{\bibfnamefont{D.}~\bibnamefont{Soriano}},
  \bibinfo{author}{\bibfnamefont{E.}~\bibnamefont{Navarro-Moratalla}},
  \bibinfo{author}{\bibfnamefont{K.}~\bibnamefont{Watanabe}},
  \bibinfo{author}{\bibfnamefont{T.}~\bibnamefont{Taniguchi}},
  \bibinfo{author}{\bibfnamefont{S.}~\bibnamefont{Manni}},
  \bibinfo{author}{\bibfnamefont{P.}~\bibnamefont{Canfield}},
  \bibinfo{author}{\bibfnamefont{J.}~\bibnamefont{Fern{\'a}ndez~Rossier}},
  \bibnamefont{et~al.}, \bibinfo{journal}{Science}
  \textbf{\bibinfo{volume}{360}}, \bibinfo{pages}{1218} (\bibinfo{year}{2018}).

\bibitem[{\citenamefont{Kresse and Hafner}(1993)}]{Kresse93}
\bibinfo{author}{\bibfnamefont{G.}~\bibnamefont{Kresse}} \bibnamefont{and}
  \bibinfo{author}{\bibfnamefont{J.}~\bibnamefont{Hafner}},
  \bibinfo{journal}{Physical Review B} \textbf{\bibinfo{volume}{47}},
  \bibinfo{pages}{558} (\bibinfo{year}{1993}).

\bibitem[{\citenamefont{Kresse and Furthm\"uller}(1996)}]{Kresse96}
\bibinfo{author}{\bibfnamefont{G.}~\bibnamefont{Kresse}} \bibnamefont{and}
  \bibinfo{author}{\bibfnamefont{J.}~\bibnamefont{Furthm\"uller}},
  \bibinfo{journal}{Physical Review B} \textbf{\bibinfo{volume}{54}},
  \bibinfo{pages}{11169} (\bibinfo{year}{1996}).

\bibitem[{\citenamefont{Perdew et~al.}(1996)\citenamefont{Perdew, Burke, and
  Ernzerhof}}]{PBE}
\bibinfo{author}{\bibfnamefont{J.~P.} \bibnamefont{Perdew}},
  \bibinfo{author}{\bibfnamefont{K.}~\bibnamefont{Burke}}, \bibnamefont{and}
  \bibinfo{author}{\bibfnamefont{M.}~\bibnamefont{Ernzerhof}},
  \bibinfo{journal}{Physical Review Letters} \textbf{\bibinfo{volume}{77}},
  \bibinfo{pages}{3865} (\bibinfo{year}{1996}).

\bibitem[{\citenamefont{Dudarev et~al.}(1998)\citenamefont{Dudarev, Botton,
  Savrasov, Humphreys, and Sutton}}]{DFTU}
\bibinfo{author}{\bibfnamefont{S.~L.} \bibnamefont{Dudarev}},
  \bibinfo{author}{\bibfnamefont{G.~A.} \bibnamefont{Botton}},
  \bibinfo{author}{\bibfnamefont{S.~Y.} \bibnamefont{Savrasov}},
  \bibinfo{author}{\bibfnamefont{C.~J.} \bibnamefont{Humphreys}},
  \bibnamefont{and} \bibinfo{author}{\bibfnamefont{A.~P.}
  \bibnamefont{Sutton}}, \bibinfo{journal}{Physical Review B}
  \textbf{\bibinfo{volume}{57}}, \bibinfo{pages}{1505} (\bibinfo{year}{1998}).

\bibitem[{\citenamefont{Pizzochero
  et~al.}(2019{\natexlab{b}})\citenamefont{Pizzochero, Yadav, and
  Yazyev}}]{Pizzochero19}
\bibinfo{author}{\bibfnamefont{M.}~\bibnamefont{Pizzochero}},
  \bibinfo{author}{\bibfnamefont{R.}~\bibnamefont{Yadav}}, \bibnamefont{and}
  \bibinfo{author}{\bibfnamefont{O.~V.} \bibnamefont{Yazyev}},
  \bibinfo{journal}{arXiv:1911.12150}  (\bibinfo{year}{2019}{\natexlab{b}}).

\bibitem[{\citenamefont{Torelli and Olsen}(2018)}]{Torelli18}
\bibinfo{author}{\bibfnamefont{D.}~\bibnamefont{Torelli}} \bibnamefont{and}
  \bibinfo{author}{\bibfnamefont{T.}~\bibnamefont{Olsen}}, \bibinfo{journal}{2D
  Materials} \textbf{\bibinfo{volume}{6}}, \bibinfo{pages}{015028}
  (\bibinfo{year}{2018}).

\bibitem[{\citenamefont{Xiang et~al.}(2011)\citenamefont{Xiang, Kan, Wei,
  Whangbo, and Gong}}]{Xiang11}
\bibinfo{author}{\bibfnamefont{H.~J.} \bibnamefont{Xiang}},
  \bibinfo{author}{\bibfnamefont{E.~J.} \bibnamefont{Kan}},
  \bibinfo{author}{\bibfnamefont{S.-H.} \bibnamefont{Wei}},
  \bibinfo{author}{\bibfnamefont{M.-H.} \bibnamefont{Whangbo}},
  \bibnamefont{and} \bibinfo{author}{\bibfnamefont{X.~G.} \bibnamefont{Gong}},
  \bibinfo{journal}{Physical Review B} \textbf{\bibinfo{volume}{84}},
  \bibinfo{pages}{224429} (\bibinfo{year}{2011}).

\bibitem[{\citenamefont{Xiang et~al.}(2013)\citenamefont{Xiang, Lee, Koo, Gong,
  and Whangbo}}]{Xiang13}
\bibinfo{author}{\bibfnamefont{H.}~\bibnamefont{Xiang}},
  \bibinfo{author}{\bibfnamefont{C.}~\bibnamefont{Lee}},
  \bibinfo{author}{\bibfnamefont{H.-J.} \bibnamefont{Koo}},
  \bibinfo{author}{\bibfnamefont{X.}~\bibnamefont{Gong}}, \bibnamefont{and}
  \bibinfo{author}{\bibfnamefont{M.-H.} \bibnamefont{Whangbo}},
  \bibinfo{journal}{Dalton Transactions} \textbf{\bibinfo{volume}{42}},
  \bibinfo{pages}{823} (\bibinfo{year}{2013}).

\bibitem[{\citenamefont{Kitaev}(2006)}]{Kitaev}
\bibinfo{author}{\bibfnamefont{A.}~\bibnamefont{Kitaev}},
  \bibinfo{journal}{Annals of Physics} \textbf{\bibinfo{volume}{321}},
  \bibinfo{pages}{2 } (\bibinfo{year}{2006}).

\bibitem[{\citenamefont{Yadav et~al.}(2018)\citenamefont{Yadav, Rachel, Hozoi,
  van~den Brink, and Jackeli}}]{Yadav18b}
\bibinfo{author}{\bibfnamefont{R.}~\bibnamefont{Yadav}},
  \bibinfo{author}{\bibfnamefont{S.}~\bibnamefont{Rachel}},
  \bibinfo{author}{\bibfnamefont{L.}~\bibnamefont{Hozoi}},
  \bibinfo{author}{\bibfnamefont{J.}~\bibnamefont{van~den Brink}},
  \bibnamefont{and} \bibinfo{author}{\bibfnamefont{G.}~\bibnamefont{Jackeli}},
  \bibinfo{journal}{Physical Review B} \textbf{\bibinfo{volume}{98}},
  \bibinfo{pages}{121107} (\bibinfo{year}{2018}).

\bibitem[{\citenamefont{Goodenough}(1958)}]{Goodenough58}
\bibinfo{author}{\bibfnamefont{J.~B.} \bibnamefont{Goodenough}},
  \bibinfo{journal}{Journal of Physics and Chemistry of Solids}
  \textbf{\bibinfo{volume}{6}}, \bibinfo{pages}{287 } (\bibinfo{year}{1958}).

\bibitem[{\citenamefont{Kanamori}(1959)}]{Kanamori59}
\bibinfo{author}{\bibfnamefont{J.}~\bibnamefont{Kanamori}},
  \bibinfo{journal}{Journal of Physics and Chemistry of Solids}
  \textbf{\bibinfo{volume}{10}}, \bibinfo{pages}{87 } (\bibinfo{year}{1959}).

\end{thebibliography}

\end{document}